\documentstyle[11pt,aaspp4]{article}

\begin{document} 

\hyphenation{PSPC}
\hyphenation{X-ray}

\slugcomment{To Appear in ApJ January 10, 1997}

\title{The Nature of the X-Ray Emission and the Mass Distributions \\
in Two Early-Type Galaxies}

\author{David A. Buote\altaffilmark{1} and Claude R. Canizares}

\affil{Department of Physics and Center for Space Research 37-241, \\
Massachusetts Institute of Technology \\ 77 Massachusetts Avenue,
Cambridge, MA 02139, \\ buote@ast.cam.ac.uk, crc@space.mit.edu}

\altaffiltext{1}{Present Address: Institute of Astronomy, Madingley
Road, Cambridge CB3 0HA, UK}

\begin{abstract}
We present spectral analysis of $ASCA$ observations of the early-type
galaxies NGC 720 (E4) and NGC 1332 (E7/S0) with emphasis on
constraining the relative contribution to the X-ray emission from hot
gas and the integrated emission from X-ray
binaries. Single-temperature spectral models yield poor fits to the
spectrum ($\chi^2_{red}\sim 3$) over the $\sim 0.5 - 5$ keV energy
range. Two-temperature models significantly improve the spectral fits
($\chi^2_{red}\sim 1.5$) and have soft-component temperatures and
sub-solar abundances consistent with previous $ROSAT$
single-temperature models ($T_{soft}\sim 0.6$ keV, abundances $\sim
0.1$) and hard-component temperatures ($T_{hard}\gtrsim 3$ keV)
consistent with those expected from a discrete component. The soft
component dominates the emission in both galaxies, especially in the
0.4 - 2.4 keV band used in previous $ROSAT$ studies: flux ratios are
$F_{hard}/F_{soft}=0.19(0.16-0.45)$ for NGC 720 $(2\sigma)$ and
$F_{hard}/F_{soft}=0.31(0.24-0.55)$ for NGC 1332 $(90\%)$. Combining
these spectral results with $ROSAT$ data we updated constraints on the
mass distributions for NGC 720 and NGC 1332. For NGC 720, which yields
the more precise constraints, the ellipticity of the intrinsic shape
of the mass is slightly reduced $(\Delta\epsilon_{mass}\approx 0.05)$
when the discrete component is added, $\epsilon_{mass}\sim 0.4-0.6$
$(90\%)$. The estimates for the total mass increase with increasing
discrete flux, and we find that models with $F_{hard}/F_{soft}=0.45$,
the $2\sigma$ upper limit, have masses that exceed by $\sim 30\% -
50\%$ those where $F_{hard}/F_{soft}=0$.
\end{abstract}


\section{Introduction}

It is well known that the hot, X-ray--emitting gas in many early-type
galaxies is one of the best probes of the mass distributions in these
systems (e.g., Binney \& Tremaine 1987; Fabbiano 1989). However, it is
also well known that the X-ray emission is of a more complex origin
than a single-phase isothermal hot gas in hydrostatic equilibrium
(e.g., Canizares, Fabbiano, \& Trinchieri 1987; Kim, Fabbiano, \&
Trinchieri 1992; Pellegrini \& Fabbiano 1994; Eskridge, Fabbiano, \&
Kim 1995). Although the relative importance of cooling flows and
supernovae winds in the hot gas has received much attention and
remains a controversial topic (e.g., Thomas 1986; Canizares et
al. 1987; Loewenstein \& Mathews 1987; David, Forman, \& Jones 1990;
Ciotti et al. 1991; Rangarajan 1995), the contribution to the total
X-ray emission from the integrated emission from X-ray binaries (e.g.,
Canizares et. al.  1987; Kim et al. 1992), and in particular, the
effect of such a discrete component on determinations of the mass
distributions of early-type galaxies, has received comparatively less
attention

We have previously analyzed in detail the mass distributions of the
early-type galaxies NGC 720 (E4) and NGC 1332 (E7) using X-ray data
from the $ROSAT$ (Tr\"{u}mper 1983) satellite (Buote \& Canizares
1994, 1996a, 1996c; hereafter BCa, BCb, and BCc). The relatively large
ratio of X-ray to blue-band optical luminosity $(L_X/L_B)$ for NGC 720
suggests that the hot gas dominates the X-ray emission (Canizares et
al. 1987; Kim et al. 1992), though NGC 1332, which has a smaller value
of $L_X/L_B$, may contain a significant fraction of discrete emission.
By assuming the X-rays are due only to hot gas in hydrostatic
equilibrium, we constrained the intrinsic shapes of the gravitating
matter distributions to have ellipticities $\epsilon_{mass}\sim
0.5-0.7$ for NGC 720 and $\epsilon_{mass}\sim 0.3-0.8$ for NGC 1332 at
the estimated 90\% confidence levels; the total masses of both
galaxies were $\sim 10^{12}M_{\sun}$ assuming an isothermal gas.  For
NGC 1332, however, we (in BCb) also considered a model for the
emission from discrete sources which was spatially distributed like
the optical light. We found that the lower limit on $\epsilon_{mass}$
decreased when the relative fraction of discrete emission to hot-gas
emission was increased; however, the derived total masses
increased. Unfortunately, the $ROSAT$ Position Sensitive Proportional
Counter (PSPC) data had insufficient spectral resolution to place
useful constraints on the relative contribution to the emission from a
discrete component and hot-gas component.

In this paper we present spectral analysis of data for NGC 720 and NGC
1332 obtained by the $ASCA$ satellite (Tanaka, Inoue, \& Holt
1994). With its superior spectral resolution over $ROSAT$, and its
sensitivity to energies above 3 keV, the $ASCA$ data allow the
relative fraction of the discrete emission and hot-gas emission to be
measured much more precisely than in the previous studies. We combine
these new results with the previous $ROSAT$ data to obtain updated
constraints on the mass distributions in these galaxies. In \S
\ref{obs} we describe the observations and reduction of the $ASCA$
data; in \S \ref{spec} we present the spectral analysis; in \S
\ref{mass} we update the mass distributions using both the $ASCA$ and
$ROSAT$ data; finally, in \S \ref{conc} we present our conclusions.

\section{Observations and Data Analysis\label{obs}}

NGC 720 was observed on July 17-18, 1993 as part of the Performance
Verification (PV) phase for a nominal exposure of 40 ks, and NGC 1332
was observed on August 5-6, 1995 as part of the Guest Observation (GO)
phase AO3 for a nominal exposure time of 60 ks.  We focus on the data
taken with the two Solid-State Imaging Spectrometers (SIS) because of
their superior spectral resolution compared to the Gas Imaging
Spectrometers (GIS).  Moreover, our primary focus is to measure the
relative importance of a soft and hard component, as well as the
temperature and abundances of the soft ($\sim 1$ keV) component that
is due to hot gas emitting thermal bremsstrahlung.  We found that the
SIS data provided the best constraints on spectral models and that
incorporating the GIS data did not significantly improve the
constraints on the spectral model parameters.  We mention that
Matsushita et al. (1994) have found that the GIS is more effective
than the SIS for analyzing the hard ($\gtrsim 3$ keV) emission in
early-type galaxies.

Because of the size of the point spread function (PSF) of the ASCA
X-ray Telescope (XRT) (Takahashi et al. 1995), the spatial
distributions of the X-ray emission of NGC 720 and NGC 1332 are
essentially those of point sources. Considering the asymmetry of the
XRT, ongoing calibration uncertainties, and relatively low
signal-to-noise ratio, $S/N$, of the observations, detailed spatial
analysis of these point-like sources is inappropriate, and we will
concentrate our attention on the spectral properties of the X-ray
data.

For the observation of NGC 720 the SIS were configured in 4-CCD mode,
bright data mode, and high bit rate. Since NGC 720 is point-like, the
principal advantage of using 4-CCD mode is to obtain a reliable
estimate of the local background.  Fortunately, the sensitivity of
4-CCD mode had not yet degraded substantially due to radiation damage
(Dotani, Yamashita, \& Rasmussen 1995) at the time of the PV-phase
observation of NGC 720. However, because of the degradation of the SIS
at the time of AO3, the configuration for the NGC 1332 observation was
set to single-CCD mode with faint data mode and medium bit rate.

To prepare each SIS observation for spectral analysis we (1) screened
the data to create a cleaned event list, (2) determined the aperture
extraction radius which optimized $S/N$, and (3) extracted the source
and background spectra. All of the reduction procedures were
implemented using the standard XSELECT, FTOOLS, and IRAF-PROS software
packages. See Day et al. (1996) for a description of the standard
$ASCA$ data reduction and analysis procedures.

We screened the SIS data using the FTOOLS routine {\it
ascascreen}. For NGC 720, we set the minimum elevation angle ($ELV$)
from the Earth for SIS0 to $15\arcdeg$, a slightly more stringent
value than the standard $10\arcdeg$, and relaxed the SIS1 to
$ELV=5\arcdeg$. The other screening parameters were kept at their
standard values, though we did have to manually set the pixel
rejection threshold to 100; we also used an intermediate value for the
Radiation Belt Monitor threshold ($RBM=250$). For NGC 1332 we found
that the standard settings for the screening parameters worked well
(with $RBM$ and pixel rejection threshold as above). The resulting
filtered exposure times for NGC 720 are 32.0 ks (SIS0) and 33.9 ks
(SIS1), and for NGC 1332 we obtained 53.6ks (SIS0) and 53.1 ks (SIS1).

Next for each SIS we determined the size of the spatial aperture that
optimized $S/N$. NGC 1332 was placed at the default position for a
point source in single-CCD mode; i.e. $\sim 5\arcmin$ along the
diagonal from where the corners of the 4 SIS chips meet on chip 1 for
the SIS0, chip 3 for the SIS1. Since the PSF is asymmetrical, it is
difficult to precisely define the center of this relatively low $S/N$
source; so we simply selected a center by eye. We then computed the
counts (with photons in all available energies) in circular apertures
of different radii. (We used circular apertures rather than an
elliptical aperture more closely resembling the shape of the PSF,
because the procedure to obtain auxiliary response files (ARFs),
$ascaarf$, which accounts for the effective area due to the XRT and
window absorption, currently performs best on circular regions.)  

In choosing a region to compute the background we had to weigh the
following considerations: the background region should be (1) far
enough away from the source so as not to be contaminated by the
galactic emission, (2) large enough to contain enough photons to
reduce statistical noise, and (3) located as close as possible to the
same region of the CCDs spanned by the source since vignetting effects
are not accounted for in the ARFs. From these considerations we
selected the background region to be an annulus ($r=41-46$ pixels, or
$r=4.4\arcmin - 4.9\arcmin)$, its center slightly shifted toward the
center of the chip with respect to the adopted galaxy center in order
to fit entirely on the chip. We found that the $S/N$ was optimized for
radii $r=25$ and 20 pixels $(2.7\arcmin, 2.1\arcmin)$ for the SIS0 and
SIS1 respectively.

Unfortunately NGC 720 was not positioned at the default position for a
point source. Rather, the center was placed very near the gap between
chips 1 \& 2 for SIS0 and chips 3 \& 0 for SIS1, almost $6\arcmin$
from where the corners of the 4 SIS chips meet on chip 1 for the SIS0,
chip 3 for the SIS1. As a result, significant flux was lost in the gap
between the chips. Again we centered by eye a circular aperture on NGC
720 and computed the counts for different radii. The aperture lay
mostly on chips 1 \& 3 for the SIS0 and SIS1 respectively, but some of
the area covered respectively chips 2 \& 0 as well. The gap was
excluded from the circles. The background was estimated from two
circular regions of radius $r=2\arcmin$, one placed near the top left
corner of chip 1 (3) and the other near the top right corner chip 2
(0) for the SIS0 (SIS1) in order to stay as far away from the galaxy
as possible while remaining on the same chips. We found that the $S/N$
was optimized for radii $r=40,32.5$ pixels $(4.3\arcmin, 3.5\arcmin)$
for the SIS0 and SIS1 respectively.

Finally, we extracted the spectra for the source and background. In
addition to the local background, we also computed the background from
the deep exposures of regions of ``blank sky'' provided from the
$ASCA$ Guest Observer Facility (GOF); we used the November 1994
versions of these observations.  An advantage of using the blank sky
observations is that their spectra may be computed from the same
region of the CCDs as the source and thus telescopic vignetting, which
is not accounted for by the ARF files, will be the same for both
source and background. Also, being deep exposures the blank sky
templates have the background level determined more precisely with
regards to photon statistics. The principal problem with the blank sky
templates is that the cosmic X-ray background varies with position on
the sky.

\section{Spectral Analysis\label{spec}}

The SIS spectrum spans the energy range 0.4 - 12 keV and, in Bright
mode, consists of 2048 Pulse Invariant (PI) bins which have been
corrected for exposure and instrument variations over the SIS. Upon
extraction, each SIS spectrum is automatically rebinned into 512 PI
channels to improve $S/N$. We further rebinned the PI channels so that
each bin had at least 40 and 50 counts for NGC 720 and NGC 1332
respectively.  We used PV-phase redistribution matrices (RMFs)
(version 0.8) provided by the $ASCA$ Team at the $ASCA$ GOF for NGC
720 and generated our own epoch-dependent (version 0.9) RMFs for NGC
1332 using the FTOOLS package $sisrsp$. The RMFs specify the channel
probability distribution for photons of a given energy; i.e. the
response matrix (RSP) is the product of the ARF (\S \ref{obs}) and the
RMF.

Because of current calibration uncertainties ($ASCA$ GOF 1996), we do
not analyze PI bins with energies $E\le 0.5$ keV. Since the X-ray
emission from early-type galaxies is generally very weak for $E\gtrsim
8$ keV (e.g., Awaki et al. 1994; Matsushita et al. 1994), we also
neglect those high-energy bins in our fits. Our motivation for this is
to reduce the systematic effects due to the background spectrum, which
dominates at those energies. Also, the contribution from foreground
and background sources will be most serious at those energies where
the $S/N$ of the galactic emission is low.

\begin{figure}[p]
\caption{  \label{fig.1t.n720} }
\raggedright

Reduced, background-subtracted SIS spectra for NGC 720 and the best-fit
single-temperature RS model (dots) with the corresponding
residuals = (model-data)/$\sigma$.

\plottwo{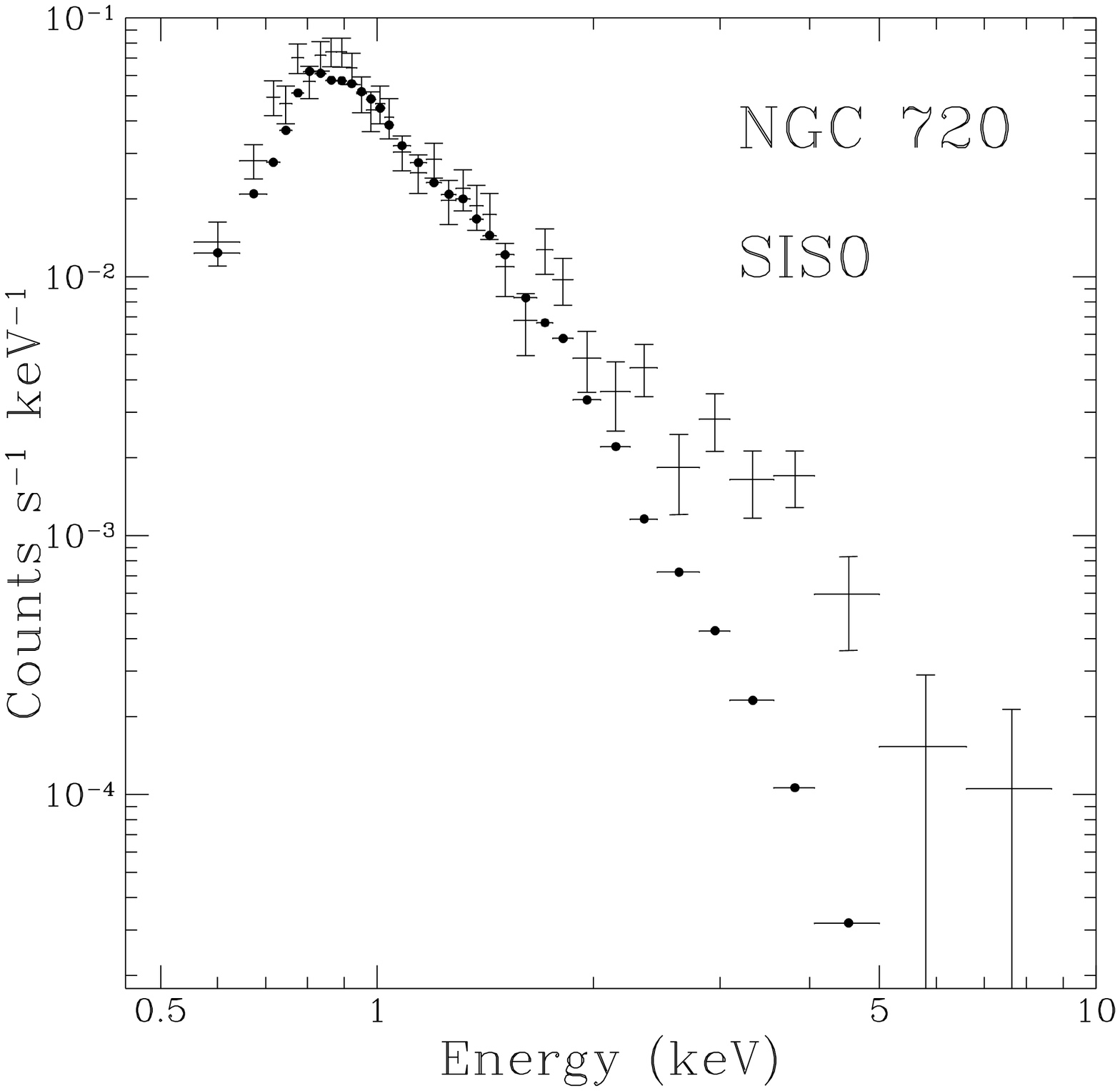}{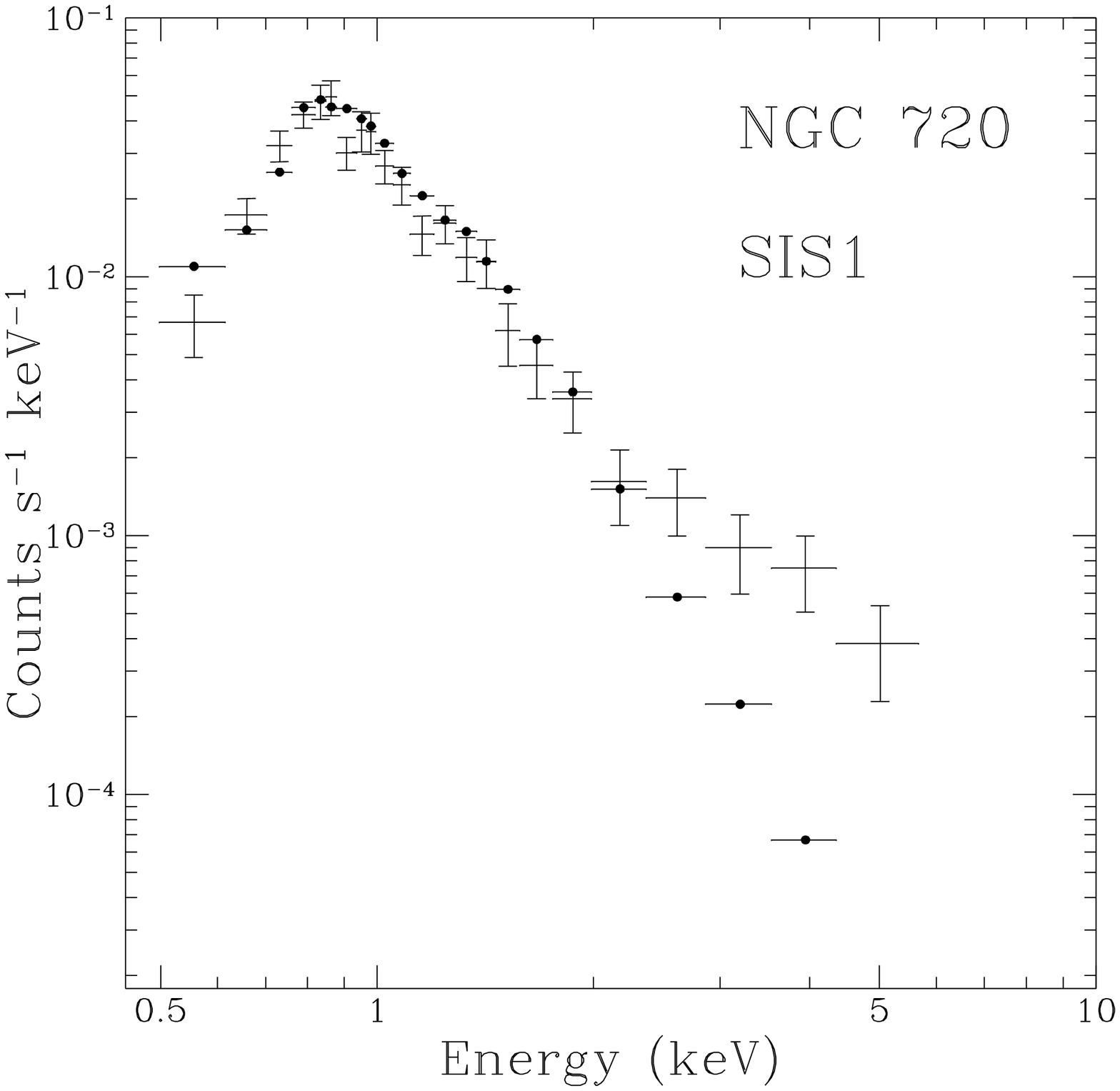}

\plottwo{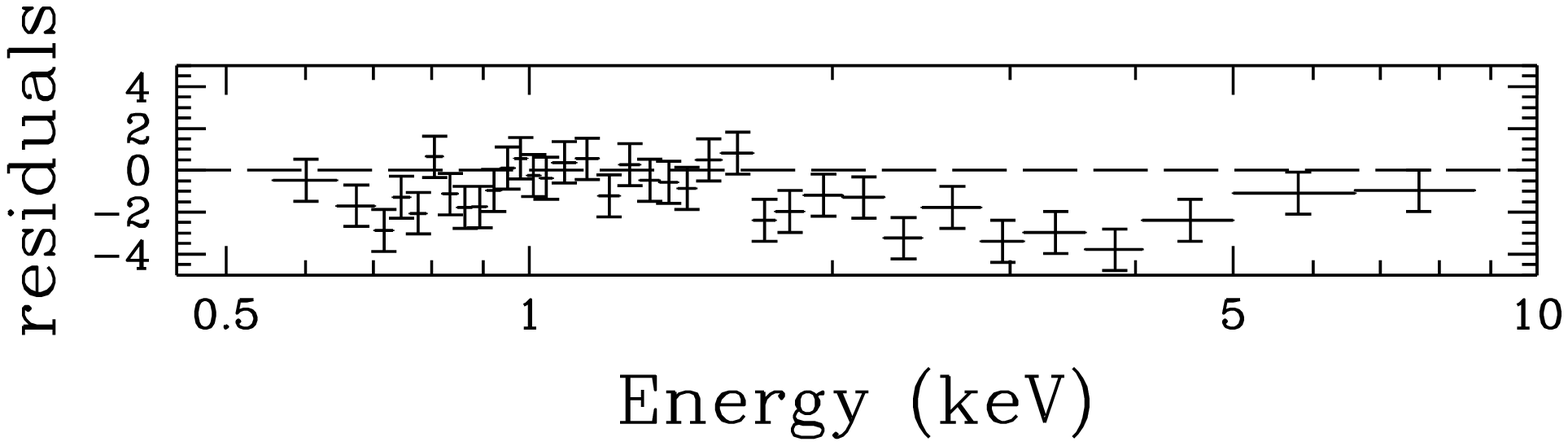}{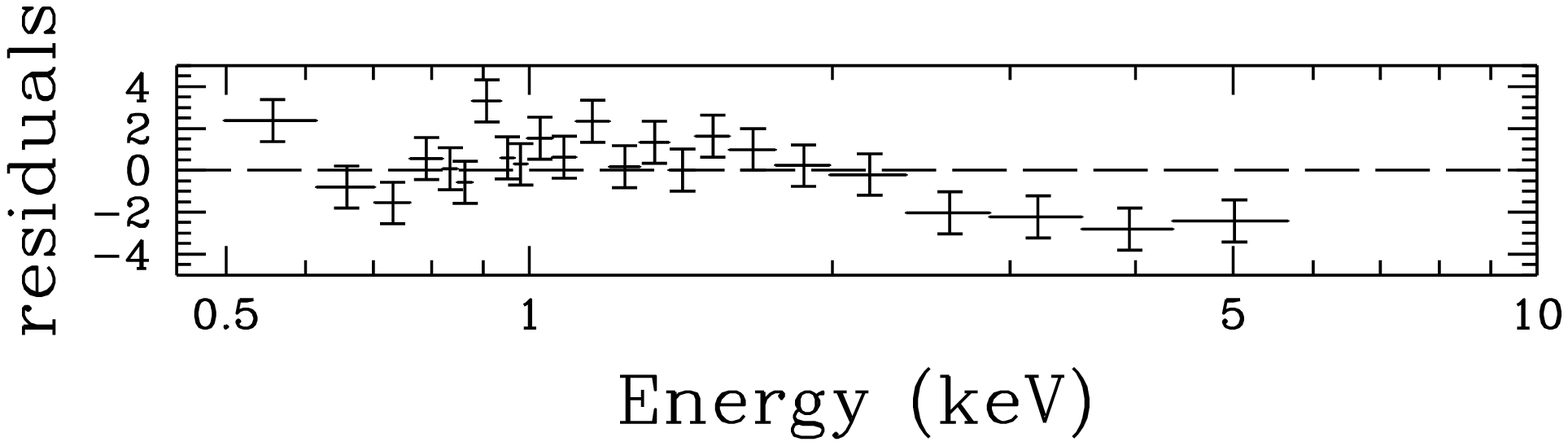}

\end{figure}

\begin{figure}[p]
\caption{  \label{fig.1t.n1332} }
\raggedright

Reduced background-subtracted SIS spectra for NGC 1332 and the
best-fit single-temperature RS model (dots) with the corresponding
residuals = (model-data)/$\sigma$.

\plottwo{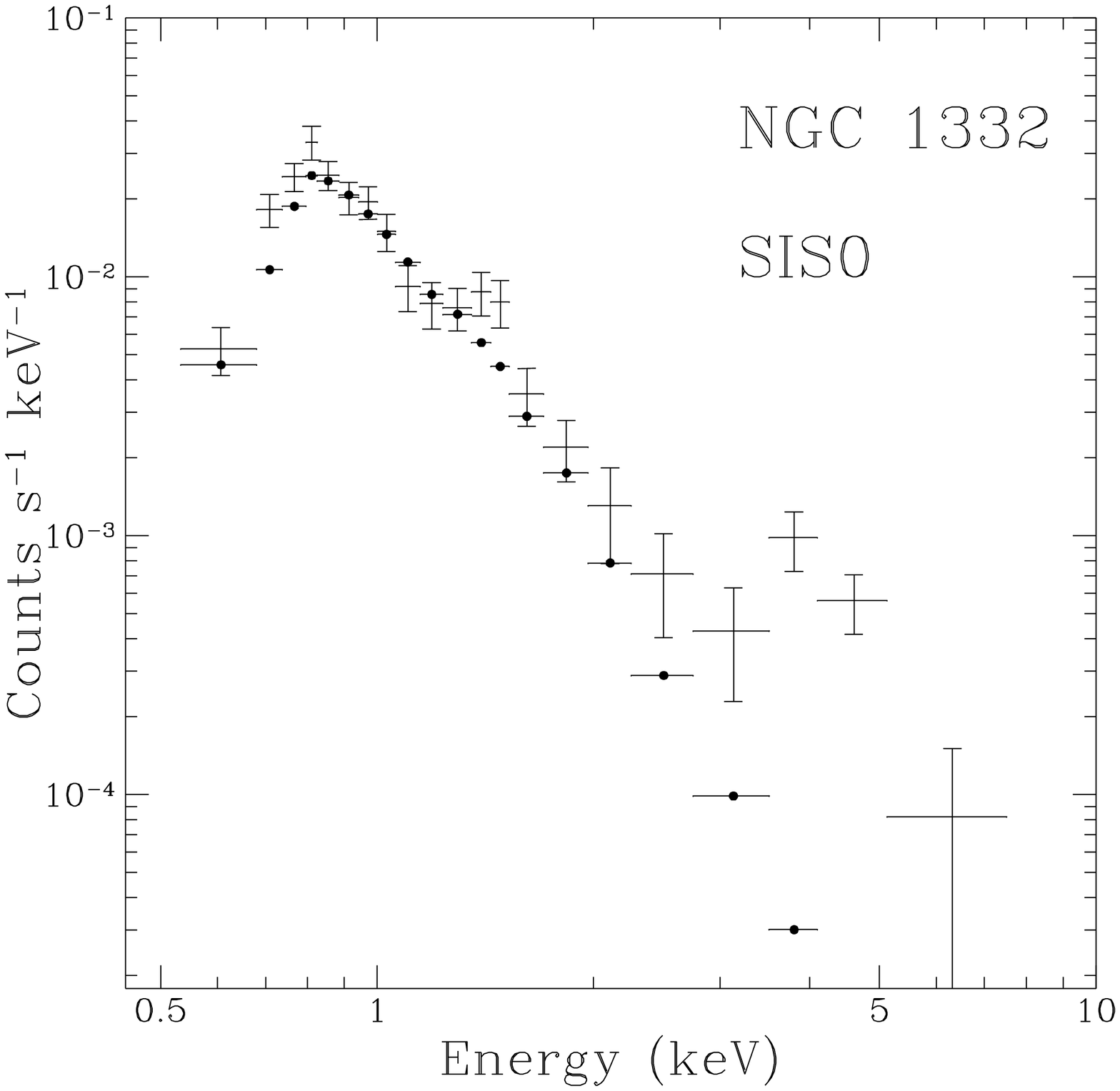}{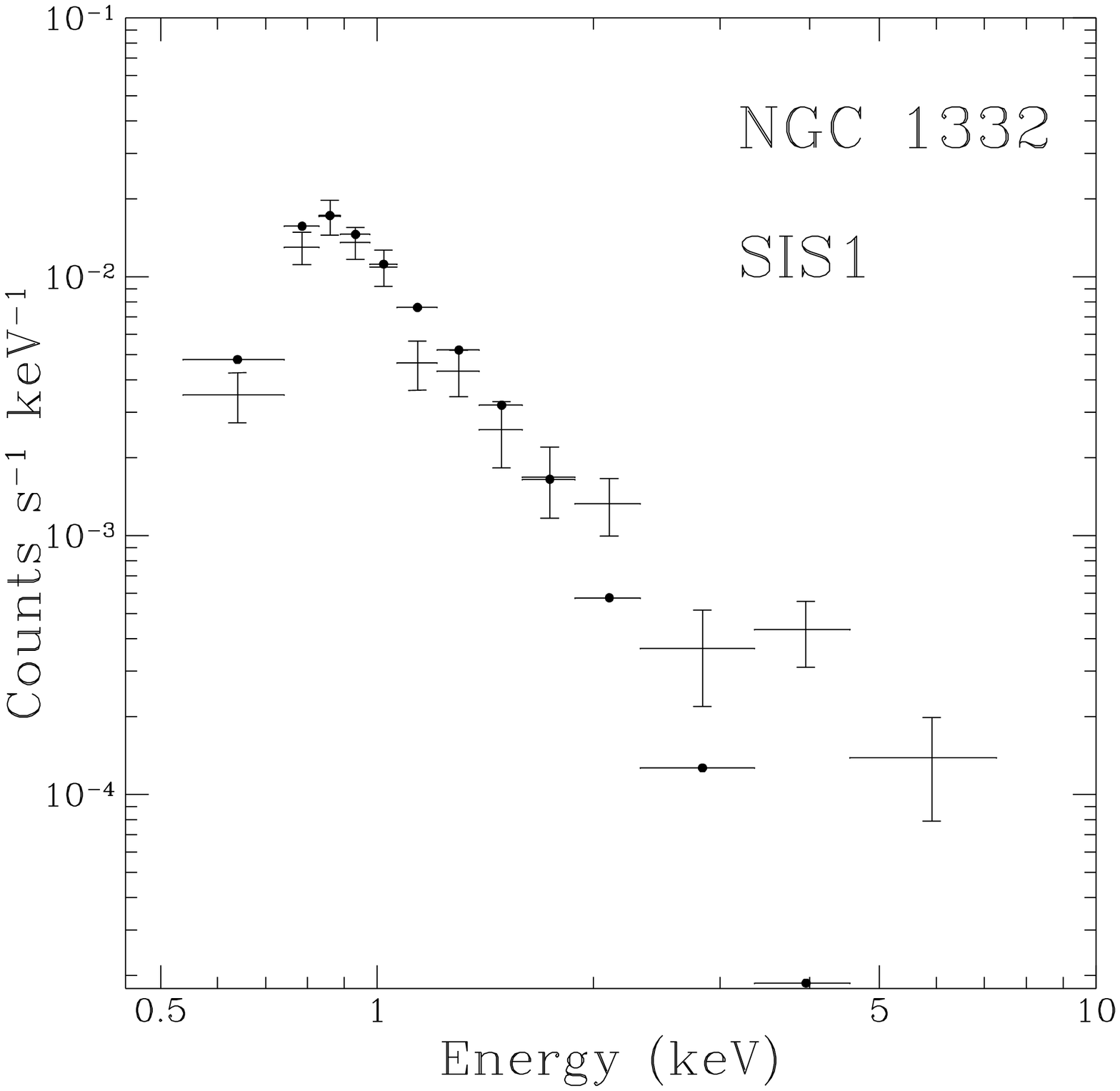}

\plottwo{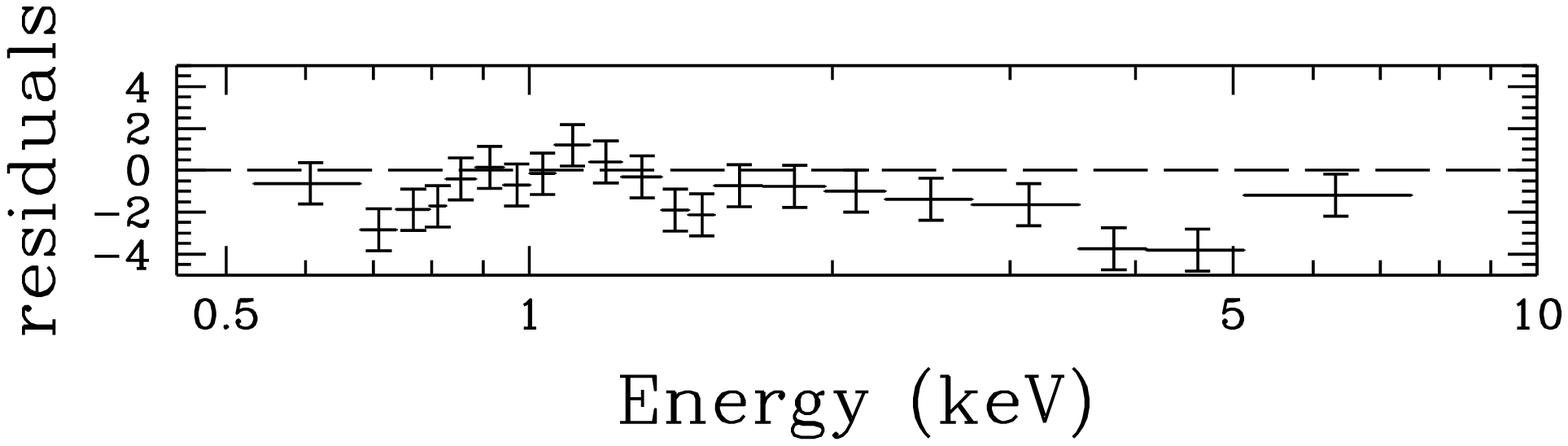}{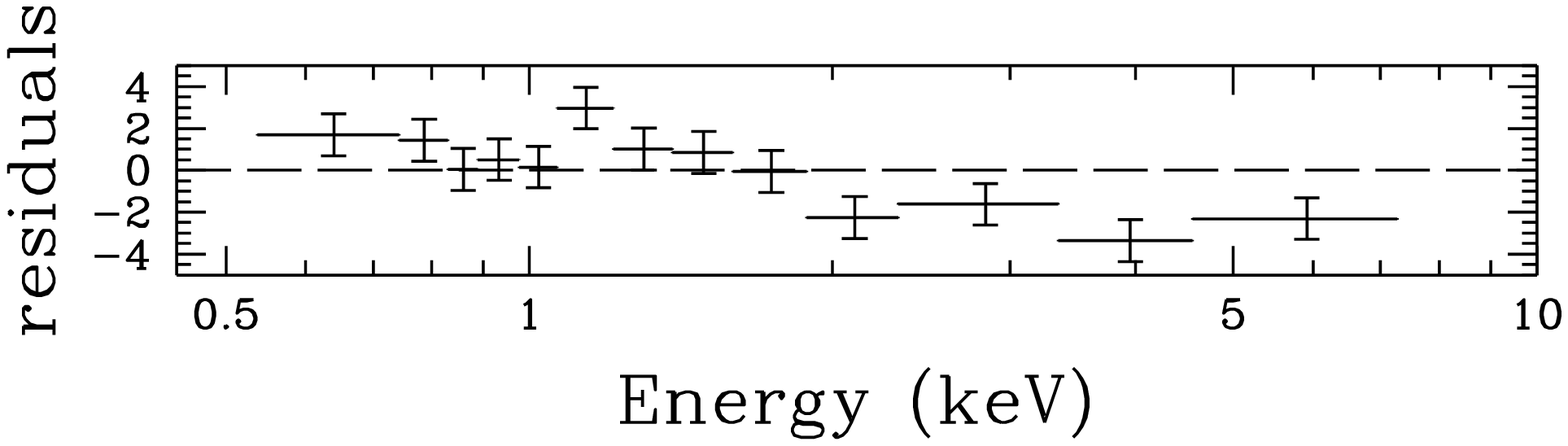}

\end{figure}

We present the background-subtracted SIS spectra for NGC 720 in Figure
\ref{fig.1t.n720} and for NGC 1332 in Figure \ref{fig.1t.n1332}. The
background-subtracted count rates are given in Table
\ref{table.flux}. The blank-sky templates are used as our standard for
the background estimates, though for comparison all of the ensuing
analysis was also performed using the local background estimates. For
N1332, the count rates are essentially consistent for both background
estimates. For NGC 720 the blank sky templates are systematically
larger than the local values by $\sim 20\%$. However, the results from
spectral fitting are quite consistent when either the blank sky
template or local background is used.

As the purpose of this investigation is to understand the relative
contribution to the X-ray emission from hot gas and discrete sources,
we first fitted the spectra to thermal models having a single
temperature. (BCc and BCb have shown that the ROSAT PSPC data rules
out single-component discrete models for NGC 720 and NGC 1332). Then
we fitted two-temperature models to the spectra to investigate whether
the fits are improved over single-temperature models. All of the
spectral fitting was performed with XSPEC.

We present the results of the spectral fits in Table \ref{table.fits}
and Figures \ref{fig.1t.n720} and \ref{fig.1t.n1332}. The results
listed in Table \ref{table.fits} all have the Hydrogen column density,
$N_H$, fixed to the galactic value as determined by Stark et
al. (1992); $N_H=(1.4,2.2)\times 10^{20}$ cm$^{-2}$ for respectively
NGC 720 and NGC 1332. We found that in all of the fits the value of
$\chi^2_{red}$ remained essentially unchanged when $N_H$ was allowed
to vary and was always consistent with the Galactic value.

The count rates for the SIS0 exceed those of the SIS1 for both
galaxies because of the greater sensitivity of the SIS0. However,
because of the greater sensitivity the extraction region used for the
SIS0, and thus the flux, is slightly larger than that of the SIS1. For
NGC 720 in particular, the models generally lie slightly below the
data for the SIS0 and slightly above the data for the SIS1. The
extraction regions differ by only about $r\sim 1\arcmin$, which is
smaller than the PSF. Since the properties of the galaxies should not
change drastically over those scales this should not pose a serious
problem in the interpretation of joint fitting of the SIS0 and SIS1
data.

The single-temperature (1T) models give very poor fits for both
galaxies because they cannot account for the emission above $E\sim 3$
keV. The Raymond-Smith (RS) models, which yield the best fits, have
best-fit temperatures and abundances that are marginally consistent
with previous determinations from the ROSAT PSPC (BCa,BCb,BCc), which
probed the energy range 0.1 - 2.4 keV. We do not give confidence
levels for the 1T models because of the poor fits.

The SIS spectra compel us to consider more complex models, of which a
natural extension are two-temperature (2T) models.  We present results
for the RS models, where we required the abundances of each
temperature component to be the same in the fits. Thus, we have four
interesting parameters: $T_{soft}$, Abun, $T_{hard}$, and the relative
normalization of the two components. We express this last parameter as
$F_{hard}/F_{soft}$, the ratio of the fluxes of the two components in
the 0.4 - 2.4 keV band.

\begin{figure}[p]
\caption{  \label{fig.2t.n720} }
\raggedright

Same as Figure \ref{fig.1t.n720} except dots are the best-fit
two-temperature RS models.

\plottwo{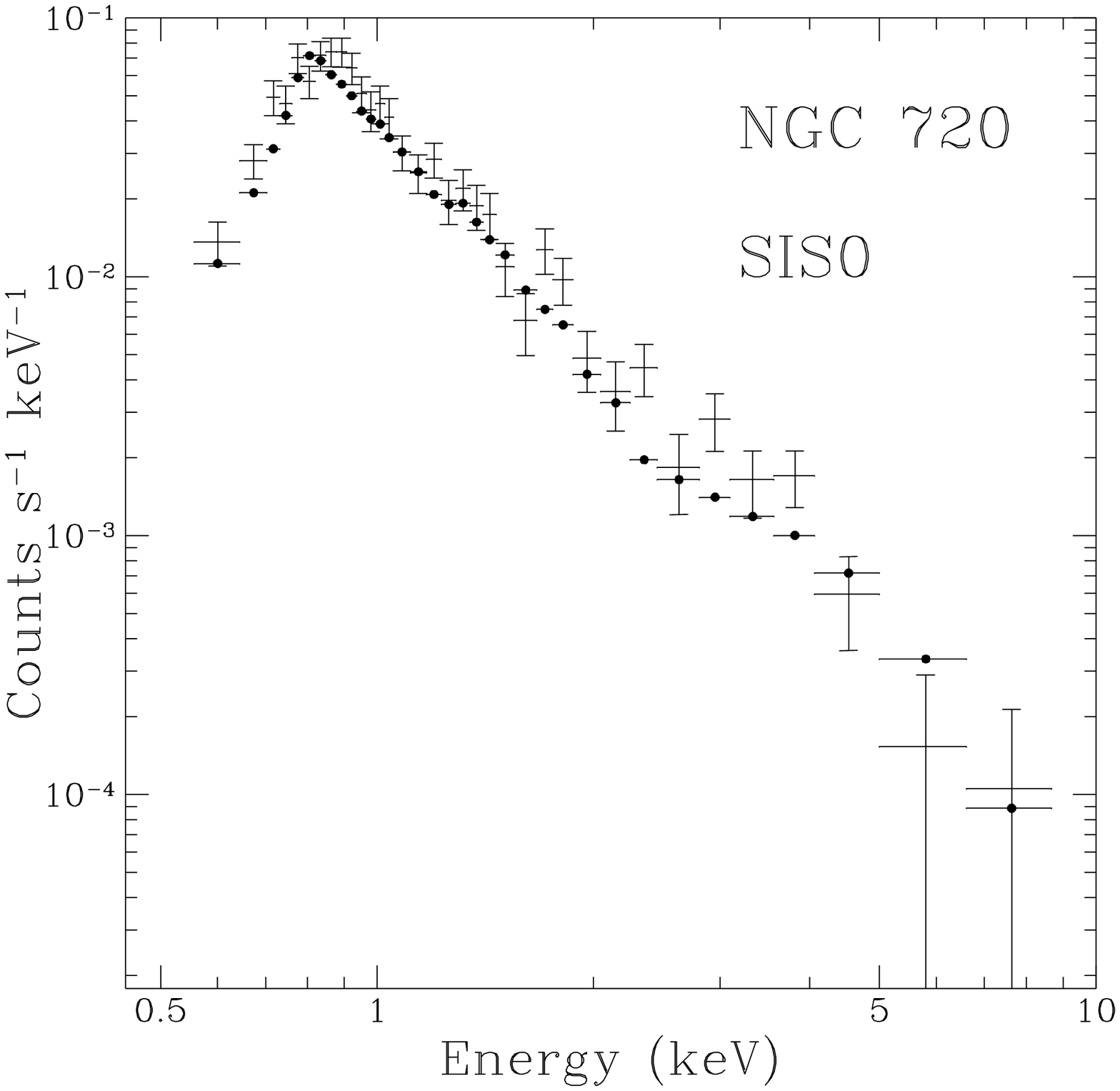}{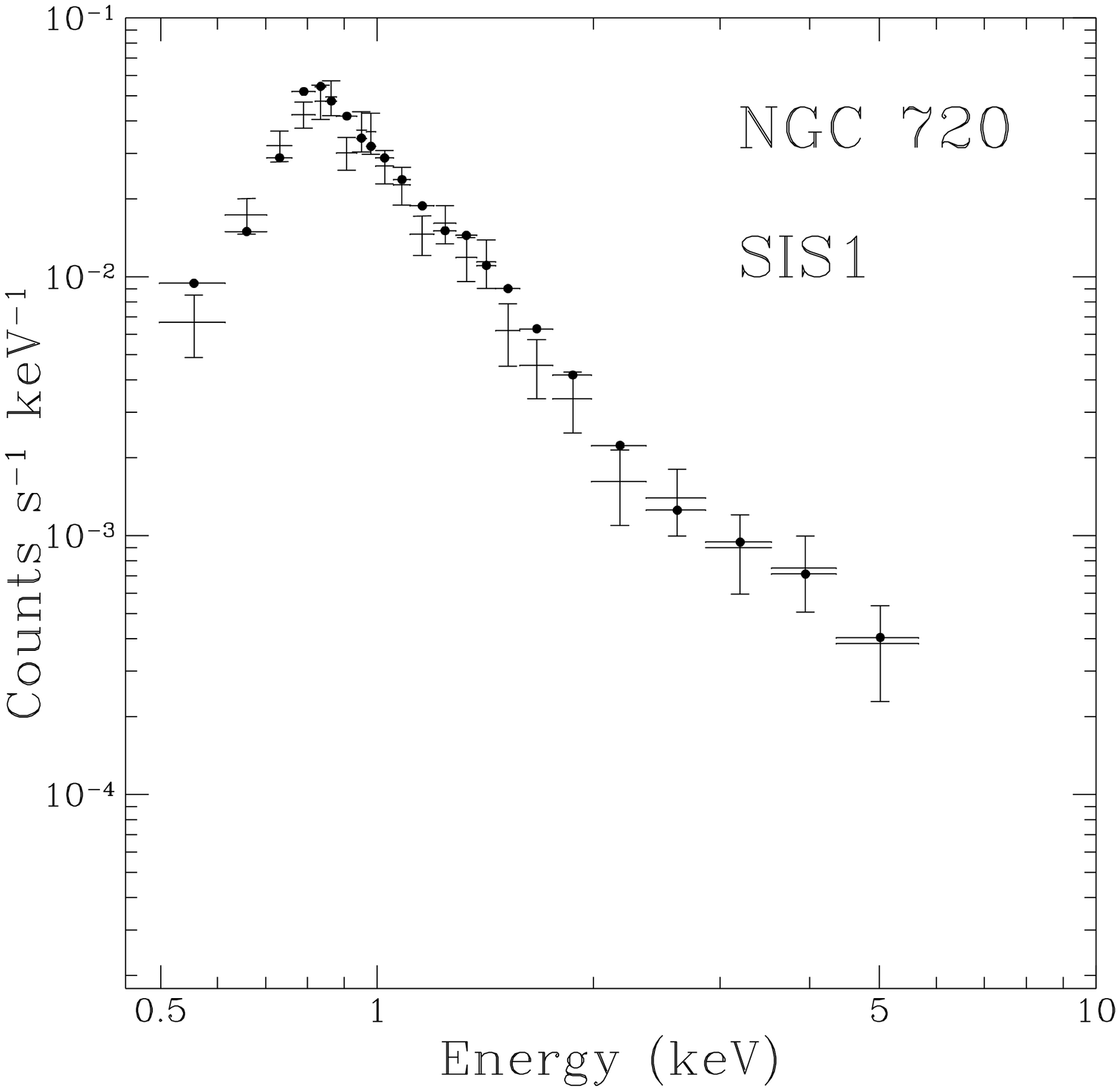}

\plottwo{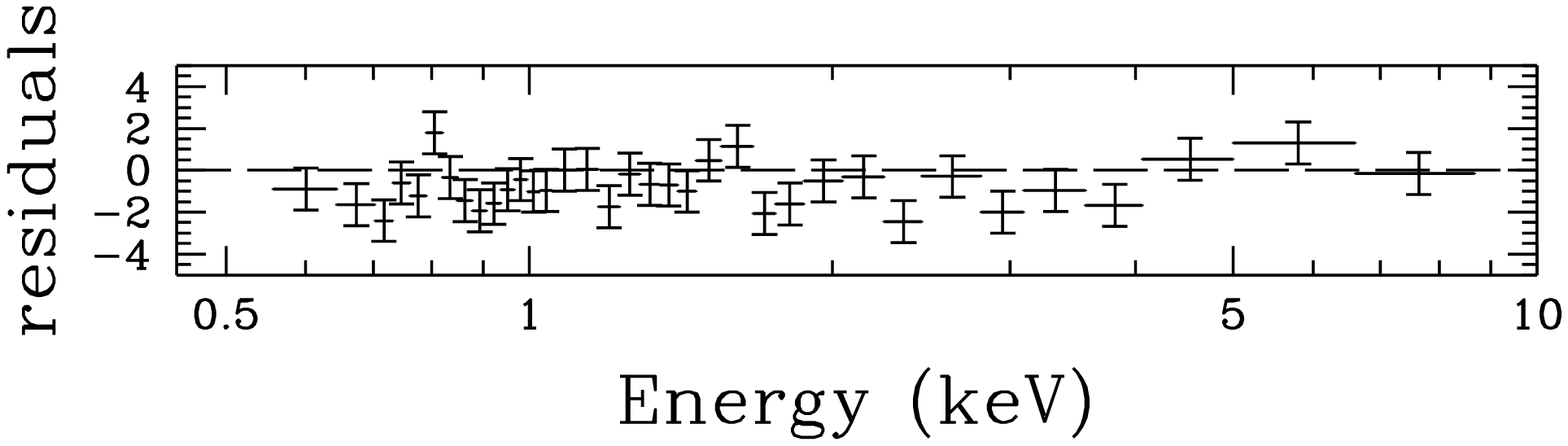}{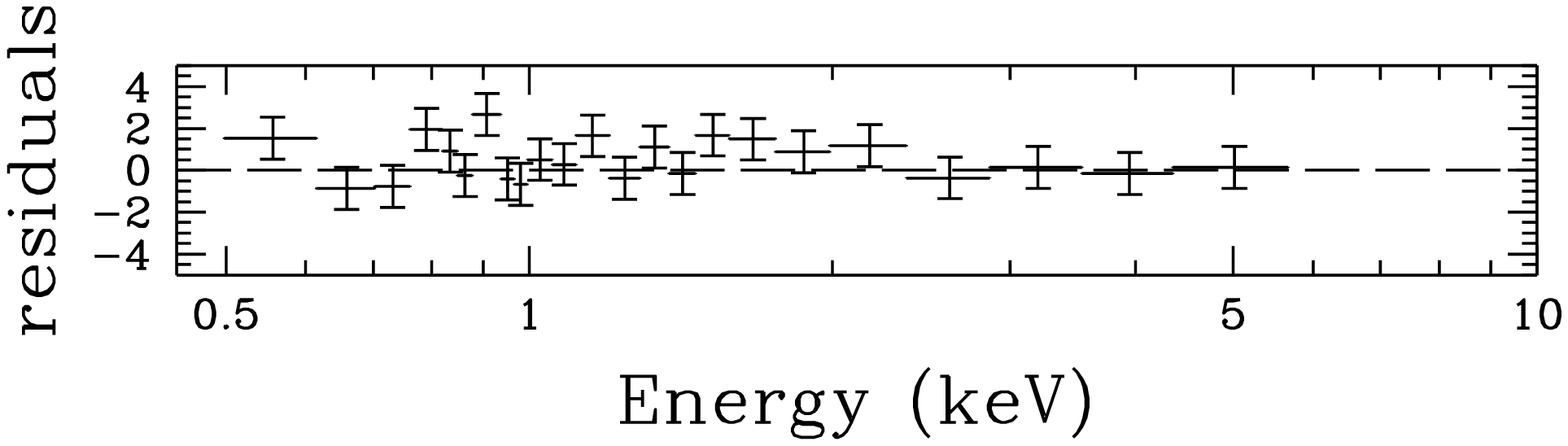}

\end{figure}

\begin{figure}[p]
\caption{  \label{fig.2t.n1332} }
\raggedright

Same as Figure \ref{fig.1t.n1332} except dots are the best-fit
two-temperature RS models.

\plottwo{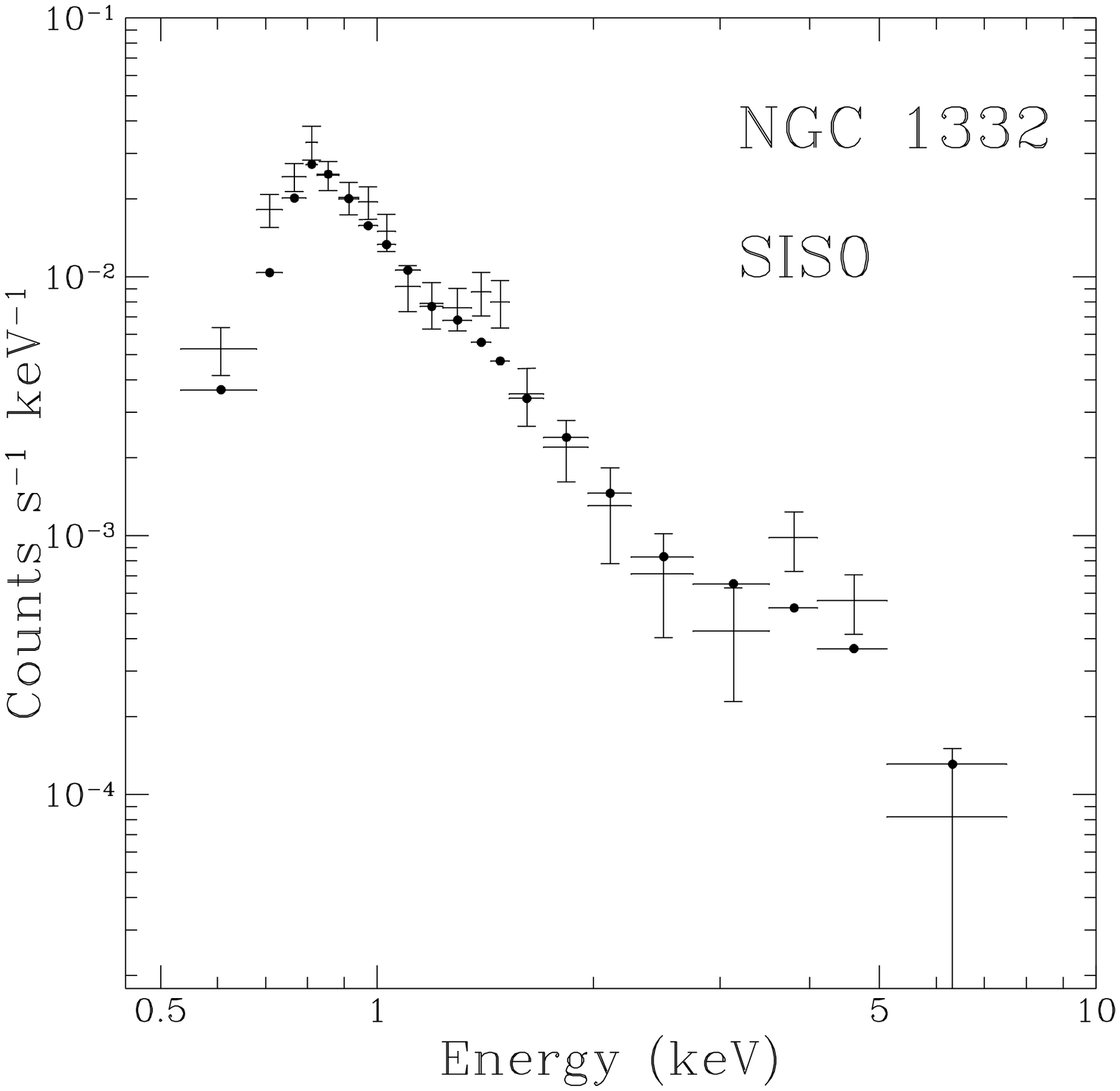}{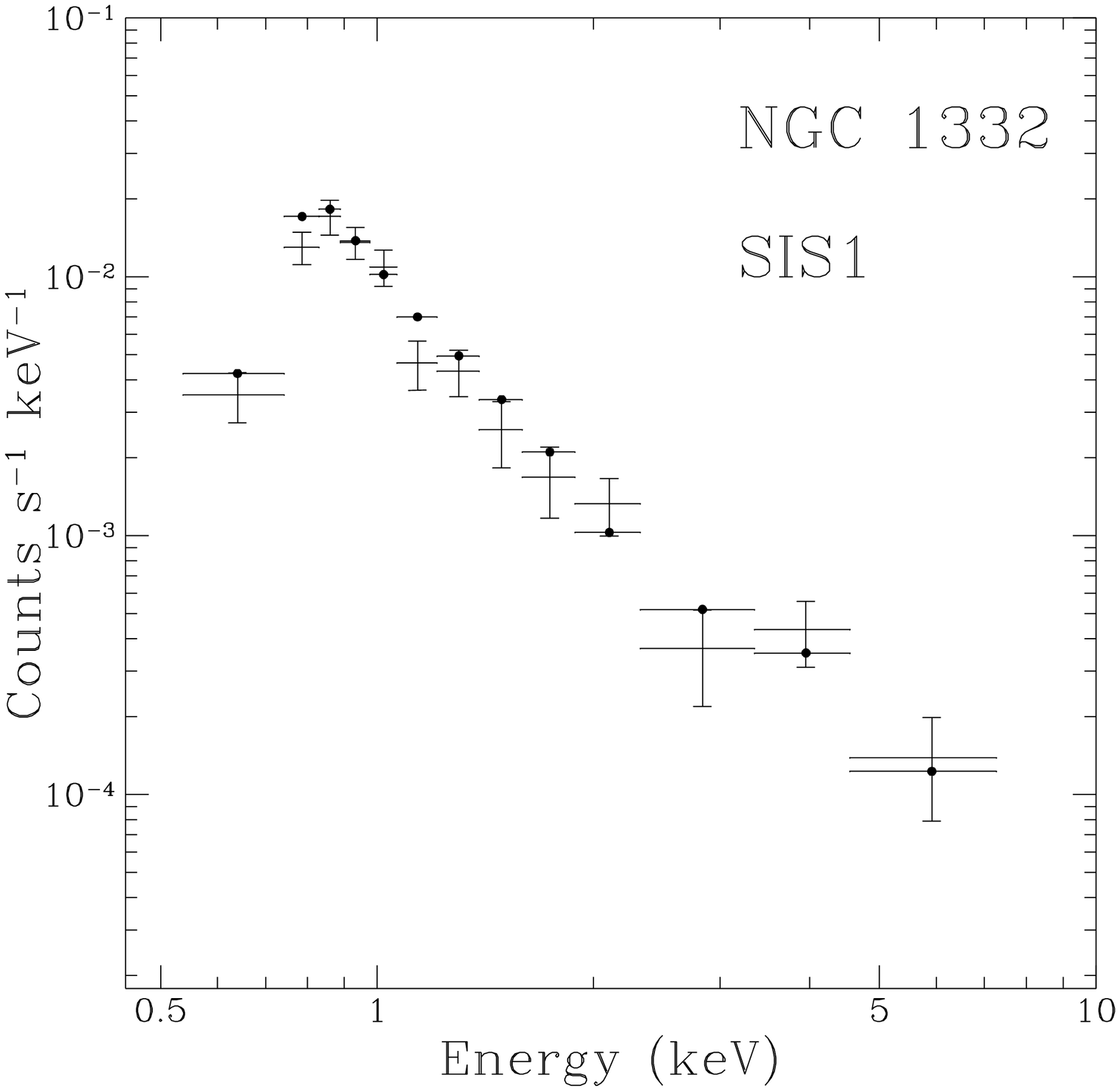}

\plottwo{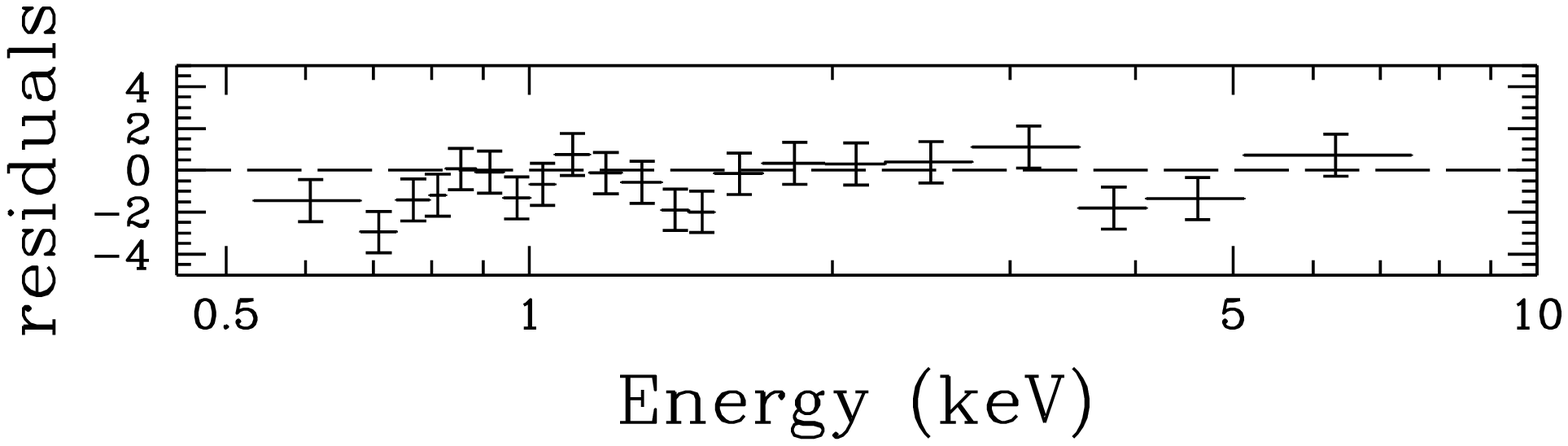}{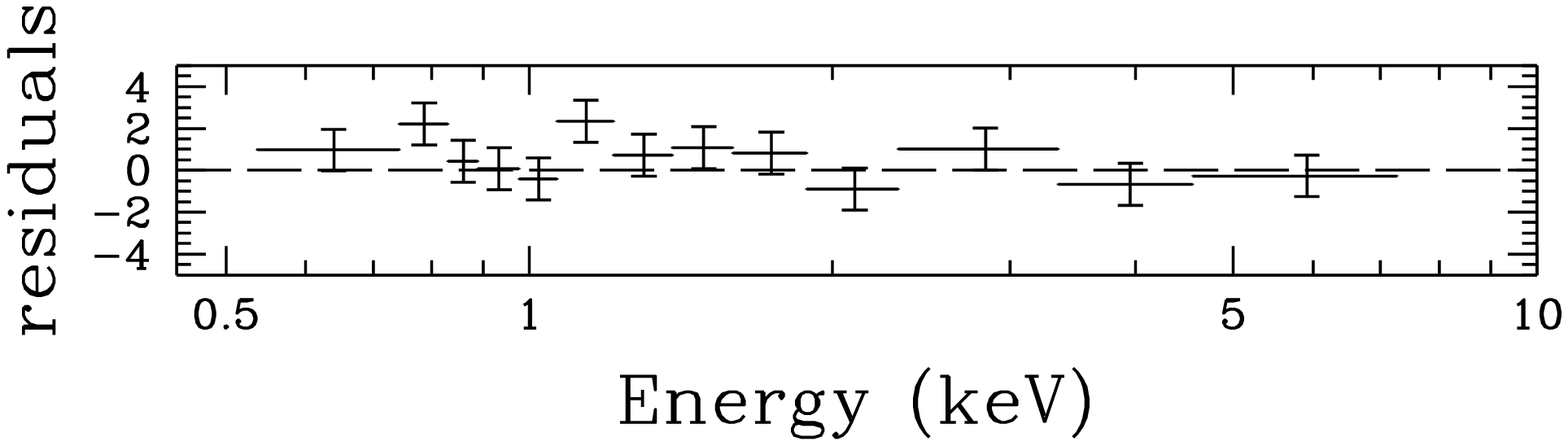}

\end{figure}

As shown in Table \ref{table.fits} and Figures \ref{fig.2t.n720} and
\ref{fig.2t.n1332}, the 2T models provide much better fits to the
spectra. Although the quality of the fits, as given by
$\chi^2_{red}\sim 1.5-1.7$, is not formally acceptable, considering
the simple assumptions of the RS models (i.e. single-phase thermal
line spectrum) over the large energy range $E\sim 0.5 - 5$ keV, the 2T
RS models would seem to be a good qualitative description for our
purposes; i.e. to determine the relative contribution to the emission
from a hard and soft component. The estimated $1\sigma$ and $2\sigma$
$(\Delta\chi^2=4.72,9.70)$ confidence levels are given for NGC 720,
while we list $1\sigma$ and $90\%$ $(\Delta\chi^2=4.72,7.78)$ levels
for NGC 1332; only 90\% confidence is given for NGC 1332 because the
lower $S/N$ does not allow constraints on the upper limit for the
abundances at $2\sigma$ which is important for determining the lower
limit on $F_{hard}/F_{soft}$.

For both galaxies the best-fit values and confidence ranges are very
similar for the parameters $T_{soft}$, Abun, and $T_{hard}$, with the
values for NGC 720 determined somewhat more precisely due to the
higher $S/N$. The values for $T_{soft}\sim 0.6 (0.5 - 0.7)$ keV and
the sub-solar abundances are consistent with previous 1T models
determined from the ROSAT PSPC for these galaxies (BCa,BCb,BCc) as
expected because of the 0.1 - 2.4 keV pass band of ROSAT.  However,
both galaxies clearly require a second, high-temperature component
($T_{hard}\gtrsim 5$ keV), that is consistent with the integrated
emission from X-ray binaries (i.e. discrete sources) in the galaxies
(Canizares et al. 1987; Kim et al. 1992). Using the {\it Einstein}
results from Canizares et al. (1987), the approximate expected 0.5 -
4.5 keV flux due to the emission from discrete sources is $\sim
1.6\times 10^{-13}$ erg cm$^{-2}$ s$^{-1}$ for NGC 720 and $\sim
1.3\times 10^{-13}$ erg cm$^{-2}$ s$^{-1}$ for NGC 1332, which are in
reasonable agreement with the fluxes in Table \ref{table.flux}.

In order to compare to previous $ROSAT$ results for NGC 720 and NGC
1332, we list in Table \ref{table.fits} the values for
$F_{hard}/F_{soft}$ in the energy band 0.4 - 2.4 keV; this band is
also the principal band of emission for hot gas with $T\sim 1$
keV.  The SIS spectra provide interesting constraints on the relative
fluxes of the two components and demonstrate that the soft component
in both galaxies, which is probably due to hot gas, dominates the
total X-ray emission in the 0.4 - 2.4 keV band.  As expected, the soft
component constitutes a larger fraction of the total 0.4 - 2.4 keV
emission in NGC 720 because of its larger ratio of X-ray to optical
blue-band luminosity, $L_X/L_B$, than NGC 1332 (Kim et al. 1992).
Over the 0.5 - 5 keV band, however, the hard component becomes more
important, though it does not dominate the total emission in either
galaxy: $F_{hard}/F_{soft} = 0.40 (0.36 - 0.59) (0.35 - 0.73)$ for
$1\sigma$ and $2\sigma$ respectively for NGC 720, and
$F_{hard}/F_{soft} = 0.65 (0.52 - 0.96) (0.52 - 1.02)$ for $1\sigma$
and 90\% respectively for NGC 1332. For clarity, the individual values
of $F_{soft}$ and $F_{hard}$ are listed in Table \ref{table.flux} for
both the 0.4 - 2.4 keV and 0.5 - 5 keV bands. The values of
$F_{hard}/F_{soft}$ are comparable both when the background is taken
from the sky templates and the local estimates.

The qualitative properties of the two spectral components for NGC 720
and NGC 1332 are similar to previous $ASCA$ analyses of a small sample
of early-type galaxies in Virgo (Awaki et al. 1994; Matsushita et
al. 1994). There are, however, some notable differences.  The inferred
lower limits on $T_{hard}$ derived in this paper generally exceed
those determined for the Virgo galaxies.  This difference would not
appear to be due to larger uncertainties for the Virgo galaxies
because their count rates are factors of 5-10 larger than for NGC 720
and NGC 1332. This result is consistent with the values of $T_{hard}$
of the Virgo galaxies being diminished because of a contribution from
the diffuse cluster emission ($T\sim 2$ keV, as suggested by
Matsushita et al. 1994), whereas NGC 1332, and especially NGC 720, do
not reside in regions of high galaxy density.  This contamination from
the diffuse cluster gas may also account for the $\sim 25\%$ larger
values of $T_{soft}$ of the Virgo galaxies in comparison to NGC 720
and NGC 1332.  

Although the abundances for NGC 1332 are not tightly constrained, NGC
720 has precisely determined RS abundances that are slightly lower
than found for the bright Virgo ellipticals NGC 4406, NGC 4472, and
NGC 4636 studied by Awaki et al. (1994) and Matsushita et al. (1994),
but are consistent with those of NGC 1404 and NGC 4374 (Loewenstein et
al. 1994).  We advise caution in interpreting the precise values of
the abundances as a result of the marginal fits of the RS models;
i.e. the residuals below 2 keV are suggestive of line emission that is
not well accounted for by the models, which would raise the abundances
to some degree. See, e.g., Loewenstein et al. (1994), for a discussion
of the implications of these low abundances for models of the origin
of the hot gas.

It is interesting to compare the values of $F_{hard}/F_{soft}$ for NGC
720 with NGC 4472 (similarly for NGC 1332 with NGC 4374) because the
galaxies have similar values of $L_X/L_B$ (Kim et. al. 1992).  Awaki
et al. (1994) obtained $F_{hard}/F_{soft}=0.09(0.04-0.15)$ (90\%
confidence) in the 0.5 - 4.5 keV band for NGC 4472 which is
substantially smaller than for NGC 720.  In contrast, the luminosities
of the hard components of NGC 4472 and NGC 720 agree to within $\sim
20\%$. Although the fluxes of both NGC 720 and NGC 4472 are dominated
by the soft component for 0.4 - 2.4 keV, the precise values, as well
as those in the broader bands, have distinctly different values.  This
suggests that although $L_X/L_B$ is a useful indicator for when the
hot gas dominates the X-ray emission, the specific flux ratio of the
hot gas and emission from discrete sources clearly depends on other
factors; e.g., the environment of the galaxies.

\section{Implications for the Mass Distributions\label{mass}}

We now consider the effects of the $ASCA$ spectral results on the mass
distributions in NGC 720 and NGC 1332.  The uncertainty associated
with the temperature profile of the hot gas in an early-type galaxy is
generally the principal limiting factor in the accuracy for
determining the mass distribution assuming hydrostatic equilibrium
(e.g., Binney \& Tremaine 1987; Fabbiano 1989).  The spatial
resolution of $ASCA$ is inadequate for measuring the radial
temperature profiles for NGC 720 and NGC 1332. Spatial (polytropic)
and spectral analysis of $ROSAT$ PSPC data indicate a nearly
isothermal gas for NGC 720 (BCa), though for NGC 1332 the temperature
profile is less certain (BCb).  The temperature profiles in other
early-type galaxies have been found to be very nearly isothermal
(e.g., Forman et. al. 1993; David et al. 1994; Trinchieri et
al. 1994; Kim \& Fabbiano 1995; Rangarajan et al. 1995).

For an isothermal gas the gravitating mass is directly proportional to
the gas temperature and thus the superior spectral resolution of
$ASCA$ should provide more robust constraints than the previous PSPC
studies. However, single-temperature spectral models fitted the PSPC
spectrum quite well for NGC 720 and NGC 1332 (BCa, BCb, BCc), whereas
the $ASCA$ data requires two components (and thus two more free
parameters). As a result, the precision of the constraints on the soft
component of the $ASCA$ data turns out to be similar to the
single-temperature values obtained with the PSPC data. Hence, the mass
distributions of NGC 720 and NGC 1332 are not substantially clarified
by the new determinations of the temperatures in this paper (see
below).

Unlike with the temperature, the $ASCA$ data give a significantly more
precise description of $F_{hard}/F_{soft}$ than in previous $ROSAT$
studies (BCb, BCc) which does affect the determination of the shape of
the mass profile and the total amount of gravitating mass.  Here we
update our previous work for NGC 720 and NGC 1332 incorporating the
new temperatures and abundances for the soft component, which we
assume is due to emission from hot gas, and the relative fluxes from
the hot gas and hard component, which we take to be directly
proportional to the optical light; for detailed explanations of our
modeling procedures see BCa and BCc for NGC 720 and BCb for NGC 1332.

\subsection{NGC 720}

We recomputed the shape of the total gravitating matter in NGC 720 and
revised our estimate of its total mass. Our analysis combined the PSPC
and High Resolution Imager (HRI) (David et al. 1995) data so that the
mass models (see below) satisfied both data sets simultaneously. The
X-ray radial profiles for the HRI and PSPC data and the ellipticities
of the HRI data are as described in BCc.  However, because our data
analysis techniques have evolved since BCa, we obtained more precise
68\% and 90\% confidence levels on the X-ray ellipticities for the
PSPC data using the Monte Carlo procedure described in BCc; the
technique for computing the ellipticities is explained in \S 2.1.2 of
BCa.  The results are listed in Table \ref{table.e0x}. For semi-major
axis $a=90\arcsec$ (which was used to constrain the mass shape in
BCa), the 90\% lower limit on the ellipticity is smaller (0.15) than
computed by BCa (0.20), though the upper limit is the same.  However,
as we show below, the derived limits on the mass shape are actually
very similar because the HRI data (BCc) gives 90\% confidence lower
limits on the ellipticity for $a\sim 90\%$ that agree with the values
determined by BCa.  In this present study we found that the lower
limits on the HRI data for $a\sim 50\arcsec - 90\arcsec$ and the new
upper limits on the PSPC data for $a\sim 60\arcsec - 75\arcsec$
provided the important constraints on the mass models.

As it is our principal goal to explore the effects of the new spectral
constraints derived in this paper, particularly with respect to
$F_{hard}/F_{soft}$, we have focused our attention on a restricted set
of models. First, we take the gas to be isothermal which is a good
assumption for NGC 720 (BCa), a more uncertain one for NGC 1332
(BCb). Second, we assume that each galaxy is viewed edge-on. Third, we
ignore the position angle twist of the X-ray isophotes in NGC 720
(BCc). Finally, we assume a spheroidal mass distribution (SMD), where
the mass density is stratified on concentric, similar spheroids; note
that if the X-ray position-angle twist is due to the projection of a
triaxial mass distribution, then the oblate and prolate spheroidal
models should bracket the possible aggregate triaxial shapes and
masses.  We take the density run to be given by a power-law with slope
-2, $\rho_{mass}\sim (a^2_c + a^2)^{-1}$, where $a_c$ is a core
radius. BCa showed that models with steeper slopes do not fit the data
as well. We have also found that the following results do not change
qualitatively if we choose models without a core, like the generalized
Hernquist (1990) models used in Buote \& Canizares (1996b).  As in our
previous studies, we fixed the semi-major axis of the boundary of the
mass spheroid to $450\arcsec$ ($43.6h^{-1}_{80}$ kpc assuming a
distance of $20h^{-1}_{80}$ Mpc for both NGC 720 and NGC 1332).

We list the results for the shape of the total gravitating mass in
Table \ref{table.shape}. For $F_{hard}/F_{soft}=0$ the ellipticity of
the mass is $\epsilon_{mass}\sim 0.45-0.65$ at 90\% confidence,
essentially as we found in BCa. The upper limit on $\epsilon_{mass}$
does not change much for $F_{hard}/F_{soft}=0.16,0.19$, although the
lower limit is reduced by $\sim 0.08$. As expected, the most extreme
behavior is observed for $F_{hard}/F_{soft}=0.45$ where the lower
limit on $\epsilon_{mass}$ falls by $\sim 0.20$, and the upper limit
by $\sim 0.10$. These results imply that for the most probable values
of $F_{hard}/F_{soft}$ obtained from the $ASCA$ data,
$\epsilon_{mass}\sim 0.4-0.6$, which would appear to be in good
agreement with the elongation of the flattest, most extended optical
isophotes for NGC 720 $(\epsilon\sim 0.45)$ and the globular cluster
system (Kissler-Patig, Richtler, \& Hilker 1995).

The total gravitating masses of these models are listed in Table
\ref{table.mass}. The masses are not a strong function of
$F_{hard}/F_{soft}$, though there is a trend of larger masses for
larger values of $F_{hard}/F_{soft}$. This results because the
discrete component, which follows the optical light, is very centrally
concentrated (Lauer et al. 1995), while the X-rays are considerably
more extended (core radius $\sim 15\arcsec$). Thus, to compensate for
a greater fraction of the emission being distributed like the optical
light, the hot-gas model must become even more extended which
translates to more mass since the asymptotic mass density slope is
fixed at -2. 

The derived masses for $F_{hard}/F_{soft}=0.16,0.19$ exceed by $\sim
15\%$ those for $F_{hard}/F_{soft}=0$, while the
$F_{hard}/F_{soft}=0.45$ masses are $\sim 30\% - 50\%$ higher than for
$F_{hard}/F_{soft}=0$. It is clear that to obtain the most precise
constraints on the masses of elliptical galaxies it is important to
consider the spatial distribution of the discrete component. Here we
have assumed it is distributed like the optical light, but it would be
better to determine this directly from the X-ray observations, which
is not possible with $ASCA$.  We mention that the derived gas mass for
NGC 720 is in good agreement with the PSPC determination of BCa,
though we now have some extra uncertainty due to the range of
$F_{hard}/F_{soft}$: $M=8.3(5.7-10.1)h^{-5/2}_{80}\times
10^{9}M_{\sun}$ $(2\sigma)$.

\subsection{NGC 1332}

Unfortunately, NGC 1332 presented some complications when
incorporating the discrete component into the formalism. Unlike NGC
720, the ``core'' of the X-ray emission of NGC 1332 is not clearly
resolved by the PSPC, and we found the width of the PSPC PSF to be
uncertain at the off-axis position for NGC 1332 (BCb); i.e. the
spatial distribution of the X-ray emission of NGC 1332 for radii
$\lesssim 15\arcsec$ is quite uncertain.  In BCb we considered mass
models having, in addition to the hot gas, a discrete component that
followed the optical light and found behavior very similar to that
described above for NGC 720.

However, we have re-examined the results of BCb and have determined
that the behavior of the hot gas + discrete models is quite sensitive
to the uncertainties in the PSF (though the hot-gas only models are
not, as shown in BCb). If the width of PSF is taken so that the core
is resolved (as in BCb), then we obtain results as before. But, when
the PSF width is increased so that the core is completely unresolved
(appropriate to the off-axis position of NGC 1332 -- see BCb),
then we find different behavior. In this case, because the X-ray core
is unresolved, the core of the mass model must decrease as the
proportion of the discrete model is increased to compensate for the
extended (albeit centrally concentrated) discrete model.

Because of this ambiguity, which requires higher spatial resolution
X-ray data (e.g., HRI) to resolve, we refrain from presenting new
shape and mass estimates for NGC 1332 analogous to NGC 720. Instead we
have computed the mass for NGC 1332 using the new temperatures and
abundances, but we use the hot-gas-only models from BCb. We use the
same type of mass model as for NGC 720 and find:
$M=(0.74-1.28)(0.62-1.46)h^{-1}_{80}\times 10^{12}M_{\sun}$ for
respectively 68\% and 90\% confidence levels for oblate models and
similarly $M=(0.53-1.02)(0.41-1.27)h^{-1}_{80}\times 10^{12}M_{\sun}$
for prolate models. As expected, these masses agree quite well with
the previous determinations in BCb; the gas masses using the new
spectral results are essentially unaltered from the results of BCb.

\section{Conclusions\label{conc}}

Our joint analysis of new $ASCA$ data and $ROSAT$ data for NGC 720 and
NGC 1332 clearly shows the existence of a hard component, but the soft
component dominates the emission in both of the galaxies.  Using the
$ASCA$ SIS data we obtained in the 0.4 - 2.4 keV band appropriate to
previous $ROSAT$ studies, $F_{hard}/F_{soft}=0.19(0.16-0.45)$ for NGC
720 $(2\sigma)$ and $F_{hard}/F_{soft}=0.31(0.24-0.55)$ for NGC 1332
$(90\%)$, where $F_{hard}/F_{soft}$ is the relative flux of the hard
and soft components.  The larger values of $F_{hard}/F_{soft}$ for NGC
720 are consistent with the notion that a larger value of $L_X/L_B$ is
related to the relative importance of the hot gas with respect to
emission from discrete sources (Kim et al. 1992).

We have explored the effects of a discrete component on the mass
distributions of NGC 720 and NGC 1332.  The estimates of the
ellipticity of the mass distribution are only slightly reduced when
taking into account a reasonable contribution from a discrete
component; the revised ellipticities taking into account the emission
from discrete sources do not systematically exceed the ellipticity of
the flattest optical isophotes $(\epsilon\sim 0.45)$.  The derived
mass of NGC 720 increases as $F_{hard}/F_{soft}$ increases.  We have
thus found that to obtain the most precise constraints on the masses
of elliptical galaxies it is important to consider the spatial
distribution of the discrete component. Here we have assumed it is
distributed like the optical light, but it would be better to
determine this directly from the X-ray observations, which is not
possible with $ASCA$, but will be with $AXAF$.

\acknowledgements

We gratefully acknowledge the kind help of K. Arnaud and K. Mukai for
patiently answering questions regarding the reduction of $ASCA$ data.
This research was supported by grants NASGW-2681 (through subcontract
SVSV2-62002 from the Smithsonian Astrophysical Observatory), and
NAG5-2921.

\vfill\eject

{

\begin{table}[p] \footnotesize
\caption{$ASCA$ Count Rates and Fluxes\label{table.flux}}
\begin{tabular}{|lcc|cc|cc|cc|} \tableline\tableline
& \multicolumn{2}{c}{Counts/s} & \multicolumn{2}{c}{$F_{soft}$} &
\multicolumn{2}{c}{$F_{hard}$} & \multicolumn{2}{c}{$F_{total}$}\\ 
& \multicolumn{2}{c}{($10^{-2}$ s$^{-1}$)} &
\multicolumn{2}{c}{($10^{-13}$ erg cm$^{-2}$ s$^{-1}$)} & 
\multicolumn{2}{c}{($10^{-13}$ erg cm$^{-2}$ s$^{-1}$)} & 
\multicolumn{2}{c}{($10^{-13}$ erg cm$^{-2}$ s$^{-1}$)}\\
Target & SIS0 & SIS1 & SOFT & BROAD & SOFT & BROAD & SOFT & BROAD\\
\tableline 
720:\\
SKY & $4.3\pm 0.14$ & $2.6\pm 0.11$ & 5.9(4.5-6.5) & 5.4(4.3-5.8) &
1.1(1.0-2.0) & 2.2(2.0-3.1) & 7.0(6.5-7.6) & 7.6(7.4-7.9) \\
LOCAL & $3.8\pm 0.16$ & $2.3\pm 0.12$ & 5.2(3.5-6.0) & 4.9(3.5-5.4) &
0.8(0.6-1.9) & 1.6(1.3-2.8) & 6.0(5.4-6.6) & 6.5(6.2-6.7)\\ \\
1332:\\
SKY & $1.6\pm 0.07$ & $0.9\pm 0.06$ & 2.7(2.0-3.2) & 2.5(2.0-2.8) &
0.8(0.8-1.1) & 1.6(1.6-2.1) & 3.5(3.1-4.0) & 4.2(4.1-4.4)\\
LOCAL & $1.6\pm 0.09$ & $0.9\pm 0.07$ & 2.7(1.8-3.3) & 2.5(1.8-2.8) &
0.9(0.9-1.3) & 1.7(1.7-2.3) & 3.5(3.1-4.2) & 4.2(4.1-4.6) \\
\tableline
\end{tabular}

\tablecomments{SKY and LOCAL refer to whether the background is
subtracted using the blank sky templates or from a portion of the
observed field. The count rates are for energies $E > 0.5$ keV. The
quantities $F_{soft}$, $F_{hard}$, and $F_{total}= F_{soft} +
F_{hard}$ refer to the fluxes determined from two-temperature models
in \S \ref{spec}. The SOFT and HARD columns refer to pass bands 0.4 -
2.4 keV and 0.5 - 5 keV respectively. The listed values of the fluxes
are for the best fit values and $2\sigma/90\%$ uncertainties for NGC
720 and NGC 1332 respectively.}

\end{table}

}

\begin{deluxetable}{lcc|cc|c|cc|cc}
\tablecaption{Results of Spectral Fits\label{table.fits}}
\tablehead{ & \multicolumn{2}{c}{$T_{soft}$} &
\multicolumn{2}{c}{Abun} & \colhead{$T_{hard}$} &
\multicolumn{2}{c}{$F_{hard}/F_{soft}$}\\
&  \multicolumn{2}{c}{(keV)} & \multicolumn{2}{c}{(\% Solar)} &
\colhead{(keV)} & \multicolumn{2}{c}{(0.4 - 2.4 keV)} &
\colhead{$\chi^2$} & \colhead{$\chi^2_{red}$}}
\startdata
\cutinhead{\Large NGC 720}
\cutinhead{1-T Models}
RS  & 0.68 & \nodata & 0.05 & \nodata & \nodata & \nodata & \nodata & 
144 & 2.7\nl
MEWE & 0.54 & \nodata & 0.00 & \nodata & \nodata & \nodata & \nodata &
259 & 4.9\nl
MEKA & 0.58 & \nodata & 0.07 & \nodata & \nodata & \nodata & \nodata &
162 & 3.1\nl
\cutinhead{2-T Models}
RS $1\sigma$ & 0.64 & 0.58 - 0.68 & 0.08 & 0.05 - 0.14 & $\ge 5.3$ & 
0.19 & 0.17 - 0.34 & 77 & 1.5\nl 
RS $2\sigma$ & 0.64 & 0.54 - 0.70 & 0.08 & 0.04 - 0.21 & $\ge 3.5$ &  
0.19 & 0.16 - 0.45 & 77 & 1.5\nl
\cutinhead{\Large NGC 1332}
\cutinhead{1-T Models}
RS  & 0.77 & \nodata & 0.10 & \nodata & \nodata & \nodata & \nodata & 
96 & 3.3\nl
MEWE & 0.50 & \nodata & 0.00 & \nodata & \nodata & \nodata & \nodata &
157 & 5.1\nl
MEKA & 0.57 & \nodata & 0.07 & \nodata & \nodata & \nodata & \nodata &
119 & 4.0\nl
\cutinhead{2-T Models}
RS $1\sigma$ & 0.63 & 0.51 - 0.69 & 0.14 & 0.06 - 0.71 & $\ge 6.7$ & 
0.31 & 0.24 - 0.52 & 48 & 1.7\nl 
RS $90\%$ & 0.63 & 0.47 - 0.70 & 0.14 & 0.05 - 3.02 & $\ge 4.8$ &  
0.31 & 0.24 - 0.55 & 48 & 1.7\nl

\enddata

\tablecomments{The notation for the spectral models (as they appear in
XSPEC) is RS for Raymond \& Smith (1977, updated to current version),
MEWE for Mewe, Gronenschild, \& van den Oord (1985), and MEKA for an
amended version of MEWE due to Kaastra (1992). Confidence ranges
for the 2-T models were determined assuming 4 interesting parameters.}

\end{deluxetable}

\begin{table}[p]
\caption{Revised PSPC Ellipticities for NGC 720\label{table.e0x}} 
\begin{tabular}{cccc} \tableline\tableline
$a$\\
(arcsec) & $\epsilon_{X}$ & 68\% & 90\%\\ \tableline 
60 & 0.10 & 0.08 - 0.19 & 0.06 - 0.23\\
75 & 0.21 & 0.15 - 0.23 & 0.12 - 0.25\\          
90 & 0.25 & 0.18 - 0.28 & 0.15 - 0.30\\
105 & 0.24 & 0.18 - 0.30 & 0.10 - 0.33\\ \tableline
\end{tabular}
\tablecomments{Ellipticity of PSPC X-ray surface brightness as a
function of semi-major axis $(a)$ computed using the Monte Carlo
procedure described in BCc.}

\end{table}

\begin{table}[p]
\caption{Updated Shape of the Gravitating Matter in NGC
720\label{table.shape}} 
\begin{tabular}{ccccc} \tableline\tableline
$F_{ds}/F_{hg}$ & \multicolumn{2}{c}{Oblate $\epsilon_{mass}$} &
\multicolumn{2}{c}{Prolate $\epsilon_{mass}$}\\
(\%) & 68\% & 90\% & 68\% & 90\%\\ \tableline 
0  & 0.57-0.63 & 0.48-0.68 & 0.52-0.58 & 0.44-0.62\\
19 & 0.48-0.60 & 0.39-0.65 & 0.46-0.57 & 0.37-0.60\\
16 & 0.50-0.60 & 0.40-0.66 & 0.48-0.57 & 0.38-0.60\\
45 & 0.40-0.52 & 0.30-0.60 & 0.39-0.50 & 0.29-0.55\\
\tableline
\end{tabular}

\tablecomments{Hydrostatic mass models assuming the gas is isothermal
following BCa. The $F_{ds}/F_{hg}=0$ values represent the $\rho\sim
r^{-2}$ mass density models of BCa revised using the HRI ellipticities
in BCc and the PSPC ellipticities in Table \ref{table.e0x}.}

\end{table}

\begin{table}[p]
\caption{Updated Mass for NGC 720\label{table.mass}}
\begin{tabular}{ccccc} \tableline\tableline

& \multicolumn{4}{c}{Mass $(10^{12}h^{-1}_{80}M_{\sun})$}\\
$F_{ds}/F_{hg}$
& \multicolumn{2}{c}{Oblate} & \multicolumn{2}{c}{Prolate}\\ 
(\%) & 68\% & 90\% & 68\% & 90\% \\ \tableline 
0  & 0.93-1.23 & 0.84-1.33 & 0.73-0.99 & 0.64-1.12\\
19 & 1.00-1.36 & 0.88-1.50 & 0.76-1.11 & 0.67-1.27\\
16 & 0.98-1.33 & 0.87-1.46 & 0.75-1.07 & 0.66-1.24\\
45 & 1.09 -1.55 & 0.95-1.76 & 0.87-1.32 & 0.74-1.54\\ 
\tableline
\end{tabular}

\tablecomments{These masses correspond to the models in Table
\ref{table.shape} and the spectral results in Tables \ref{table.flux}
and \ref{table.fits}. The confidence limits are computed for a given
value of $F_{ds}/F_{hg}$.}

\end{table}


\clearpage

\end{document}